**Title:** Virtual organelle self-coding for fluorescence imaging via adversarial learning

**Short title:** Deep learning-based method for fluorescence cell staining

**Authors and affiliations:** Thanh Nguyen[1, *], Vy Bui[1], Anh Thai[1], Van Lam[2], Christopher B. Raub[2], Lin-Ching Chang[1], and George Nehmetallah[1]

[1] Electrical Engineering and Computer Science Department, The Catholic University of America, Washington DC, 20064, USA
[2] Biomedical Engineering Department, The Catholic University of America, Washington DC, 20064, USA

Email address:
Thanh Nguyen: 32nguyen@cua.edu
Vy Bui: 01bui@cua.edu
Anh Thai: 15thai@cua.edu
Van Lam: 75lam@cua.edu
Christopher B. Raub: raubc@cua.edu
Lin-Ching Chang: changl@cua.edu
George Nehmetallah: nehmetallah@cua.edu

* Corresponding author: Thanh Nguyen: 32nguyen@cua.edu
8031 Eastern Ave, Apt 211
Silver Spring
Maryland, 20910
USA
Tel: +1(443) 808-6169

**Abstract:**

Fluorescence microscopy plays a vital role in understanding the subcellular structures of living cells. However, it requires considerable effort in sample preparation related to chemical fixation, staining, cost, and time. To reduce those factors, we present a **vir**tual **fluo**rescence staining method based on deep neural **net**works (VirFluoNet) to transform fluorescence images of molecular labels into other molecular fluorescence labels in the same field-of-view. To achieve this goal, we develop and train a conditional generative adversarial network (cGAN) to perform digital fluorescence imaging demonstrated on human osteosarcoma U2OS cell fluorescence images captured under Cell Painting staining protocol. A detailed comparative analysis is also conducted on the performance of the cGAN network between predicting fluorescence channels based on phase contrast or based on another fluorescence channel using human breast cancer MDA-MB-231 cell line as a test case. In addition, we implement a deep learning model to perform autofocusing on another human U2OS fluorescence dataset as a preprocessing step to defocus an out-focus channel in U2OS dataset. A quantitative index of image prediction error is introduced based on signal pixel-wise spatial and intensity differences with ground truth to evaluate the performance of prediction to high-complex and throughput fluorescence. This index provides a rational way to perform image segmentation on error signals and to understand the likelihood of mis-interpreting biology from the predicted image. In total, these findings contribute to the utility of deep learning image regression for fluorescence microscopy datasets of biological cells, balanced against savings of cost, time, and experimental effort. Furthermore, the approach introduced here holds promise for modeling the internal relationships between organelles and biomolecules within living cells.

**Introduction**

Microscopy techniques, particularly the family of epifluorescence modalities, are workhorses of modern cell and molecular biology requiring microscale spatial insight. Intensity- and/or phase-based microscopy techniques such as brightfield, phase contrast, differential interference contrast, digital holography, Fourier ptychography, and optical diffraction tomography [1-7], among other modalities, have the potential to visualize the subcellular structure. However, these methods depend largely on light scattering that is defined by the internal structure-based index of refraction, which lacks biomolecular specificity. Fluorescence-based techniques, on the other hand excite fluorophores which act as labels to spatially localize biological molecules and structures within cells. These imaging techniques, especially fluorescence approaches, can involve time-consuming preparation steps, costly reagents, introducing the possibility of signal bias due to photobleaching, and misinterpretation of images due to cell damage from intense ultraviolet wavelengths used to excite molecular labels. Thus, microscopy of cells is challenging due to inherent tradeoffs in sample preservation, image quality, and data acquisition time, and inherent variability between labeling experiments.

Deep convolutional neural networks (DCNNs) capture nonlinear relationships between images globally and locally, resulting in significantly improved performance for image processing tasks compared to the traditional machine learning methods. In many studies of subcellular structure, multiple biomolecules were tracked under fluorescent image modalities to gain a fully investigation on cell morphology. Applications using DCNNs in fluorescence microscopy were investigated in super-resolution [8-11], image restoration [12], image analysis [13], and histological staining [14]. In most of the cases, cells are required to be fixed and stained with multiple specific fluorophores, fluorochromes, or fluorescent dyes that increase the cost, time, and labor. Recently, DCNNs have been employed to create digital staining images by training a pair of images to transform transmitted light microscopic images into fluorescence images [15, 16], quantitative phase images into equivalent bright-field microscopy images that are histologically stained [17]. Following on the success of these results, we present a DCNN-based computational microscopy technique employing a customized conditional generative adversarial network (cGAN) to model the relationships between distinct but correlated imaging modalities. The objectives of our work are twofold. First, to provide an end-to-end ranged-autofocusing solution for fluorescence imaging; second, to generate the corresponding fluorescence images directly from other co-registered fluorescence channels. Furthermore, we performed new customized quantitative assessment of spatial and intensity error signals that lead to better evaluation of error, and identified reasons why some image channels can predict other channels with various degrees of success. The success of our work can significantly reduce the cost and the efforts in imaging preparation, data acquisition time while preserving high-grade image qualities. Additionally, our method can significantly reduce the possibility for cell damages such as phototoxicity and photobleaching from traditional cell screening. Most importantly, our proposed approach leads to many applications of using Deep learning to studies of complex protein structures and protein relationship modeling that may not be possible to achieved, such as in [15, 16].

To achieve both objectives, we trained the customized conditional generative adversarial networks (cGAN) and experimentally demonstrated their efficacy on U2OS dataset containing only F-actin images (U2OS-AF), co-registered phase contrast - fluorescent images of human breast cancer MDA-MB-231, and human osteosarcoma U2OS cell fluorescence images with Cell Painting staining protocol (U2OS-CPS), as shown in the Results and Methods Sections.

**Results**

In this section, we describe the results from all implementations in 3 categories: (1) Autofocusing – defocusing out-of-focus images; fluorescence prediction from (2) phase contrast images and (3) fluorescence images. We combined (2) and (3) into one section for comparative purposes.

**Autofocusing**

We trained a DCNN model, name "autofocusing model" (AF model) to predict focused images from blurred out-of-focus images. The model was trained and tested on F-actin channel of U2OS-AF dataset before applying the DCNN on out-of-focus images used in Fluo-Fluo 2, 3, 4 model's training and testing (see Fig. 1). The model takes a single out-of-focus and a focused image as a pair of input-output sample for training. Images in U2OS-AF were acquired from one 384-well microplate containing U2OS cells stained with phalloidin at 20x magnification, 2x binning, and 2 sites per well. To support the Fluo-Fluo models with U2OS-CPs dataset, we only chose fixed out-of-focus range [-12µm, 8µm] in U2OS-AF dataset that covers the whole range of out-of-focus levels in U2OS-CPS dataset. We do not use images located inside the [-2µm, 2µm] range from the focus plane since they appear as focused as the ground truth ($z = 0$). In facts, it is hard to distinguish where the focus planes are for the data in the range [-2µm to 2µm]. Notice that the focused image will be repeated in many input-output pairs for different out-of-focus images at different axial planes. Figure 2 shows predicted results on testing data of U2OS-AF with different out-of-focus distances with several zoom-in regions of the most out-of-focus distances. U2OS datasets will be described in Method - data preparation Section.

Autofocusing model's performance was evaluated by the mean absolute error (MAE), peak signal to noise ratio (PSNR), and structural similarity (SSIM) on 64 either predicted images or input images with corresponding ground truth fluorescence images in each z-depth in U2OS-AF (Supplementary Figure 1). In order, average scores are: **0.01**/0.012, **37.56**/35.639, and **0.924**/0.9 for MAE, PSNR, and SSIM, respectively. Scores in bold demonstrate the feature enhancement from autofocusing model to generate images more plausibility to the ground truth rather than input. Then the autofocusing model is used to perform the prediction on out-of-focus subset data chosen manually from U2OS-CPS dataset to preserve the number of data samples. There is no ground truth for the focused images in U2OS-CPS dataset, but we qualitatively evaluated the success of the network from observation based on focusing level and then used these images in training and analyzing purposes. Typical results are shown in Fig. 3.

**Phase contrast / fluorescence to fluorescence**

We trained three cGAN models (PhC-Fluo 1, 2) to generate DAPI, Factin, and Vinculin from phase contrast images, respectively and four cGAN models (Fluo-Fluo 1, 2, 3, 4) for fluorescence prediction (e.g. Vinculin) from another fluorescence image (e.g. F-actin) (see Fig. 1).

Following the training phase, the trained PhC-Fluo and Fluo-Fluo networks were blindly tested on testing data separated from training data. Figure 4 shows our results on breast cancer MDA-MB-231 corresponding PhC-Fluo 1, 2 and Fluo-Fluo 1. With the same amount of data and training process, predicted results were similar to the ground truth in the case of generating DAPI/Hoechst from PhC (PhC-Fluo 1, 1$^{st}$ row in Fig. 4) but the predicted Vinculin from phase contrast images were not similar to the ground truth (PhC-Fluo 2, 2$^{nd}$ row in Fig. 4). Meanwhile, focal adhesion vinculin location has been localized more accurately from F-actin (Fluo-Fluo 1, 4$^{th}$ row in Fig. 4). The predicted fluorescence results of these four deep learning models presented in Fig. 4 show non-robustness in using cGAN-based framework in predicting complex fluorescence structures such as F-actin and Vinculin. However, better performance was achieved from predicting Vinculin from F-actin. Figure 5 shows Vinculin prediction from PhC_Fluo 2 and Fluo-Fluo 1 models, the absolute pixel-wise error between ground truth and vinculin predicted from either (top row) phase contrast or (bottom) F-actin images. Green area is binary mask after thresholding of 50 (19% of 255 bit-depth) over absolute pixel-wise error that indicates F-actin helps predict the location of Vinculin signal slightly better than when using phase contrast images as training inputs. To evaluate Vinculin signal prediction, we used a customized matrix of performance (see Methods section).

Similarly, Fluo-Fluo 2, 3, 4 models were implemented with the same cGAN framework. The difference between these models and the Fluo-Fluo 1 is in the inputs that contain two fluorescent proteins to predict the targeted protein. We trained and tested the models on Human U2OS-CPS cell dataset. These models can be used or edited as a pre-trained models w/wo transfer learning on a completely new types of dataset, thus making the proposed technique generalizable which is very important attribute. Transforming *one* fluorescence channel into another *one* can use *one-one* pairs of channels as input-output pairs. However, the success of model training depends on the correlation of the selected pairs, i.e., strong-correlated pairs of input-output allows the model to learn a pixel-to-pixel transformation that is governed by the regularization of the network. This data-driven cross-modality transformation framework is effective because the input and output distributions share a high degree of mutual texture information, with an output probability distribution that is conditional upon the input data distribution [9]. Performing experiments to compare the model performance based on the choices of different input-output pairs would be an interesting future research topic.

In previous studies [15, 16], predicting fluorescent proteins from transmitted microscopy were carried out successfully where well-defined cellular and high-correlated proteins to inputs were fully investigated. These findings strengthen DCNN as a state-of-art method in image transformation. In the proposed work, we determined individual structure by inputting two-channel fluorescent proteins into the DCNN models e.g., [Golgi apparatus (plasma), membrane (F-actin) + nucleoli/cytoplasmic RNA] to predict [mitochondria, nucleus (Hoechst), and endoplasmic reticulum]) (See. Fig. 6). All organelles were acquired from a microscopy assay imaging system using Cell Painting staining protocol [18], with six stains imaged across five channels, revealing eight cellular components/structures. Due to the huge amount of data collected, some of the images are out-of-focus and cannot be readily used in any data-driven analysis, especially in Golgi apparatus plasma and membrane F-actin. Hence, a necessary autofocusing DCNN model (described in Autofocusing Section above) was developed to predict focused images to be used for training Fluo-Fluo x models to perform any data-driven analysis.

To measure the performance of the proposed models, we compute the mean absolute error (MAE), peak signal to noise ratio (PSNR), and structural similarity (SSIM) on 96 predicted mitochondria, nucleus (DAPI/Hoechst), and endoplasmic reticulum images and their corresponding ground truth fluorescence images in testing dataset of U2OS-CPS (Supplementary Fig. 2). The average scores are: MAE [0.0023, 0.0106, 0.0068], PSNR [48.1931, 37.7700, 41.3456], SSIM [0.9772, 0.9564, 0.9731], for DAPI/Hoechst, Endoplasmic reticulum, and Mitochondria, respectively. We also compare the feature measurements extracted from two groups of the combinations of input-output and input-ground truth, respectively, using the modified pipeline on CellProfiler [18, 19]. Figures 7 and 8 show **P**earson product-**M**oment **C**orrelation coefficient (PMC) of each feature measurement, distributed in cell, nuclei, and cytoplasm group, across sub-amount in testing images between original (5) channels and hybrid-virtual (2+3) channels and their histograms, respectively, (see Supplementary Fig. 3 on how PMC is extracted). After removing prefect correlation of 1 (measured from input [Golgi apparatus (plasma), membrane (F-actin) + nucleoli/cytoplasmic RNA] only), the histograms show highly correlation in feature measurement between the original channels and the hybrid-virtual channels and that demonstrates the reliability in using virtual channels for biological analysis. Extracted feature measurements include correlation, granularity, intensity, radius distribution, size & shape, and texture are shown in the Supplementary Methods Section and in Supplementary Fig. 4.

**Discussions**

Unlike large depth-of-field capable imaging technique such as holography, fluorescence microscopy lacks the capability for image propagation which is necessary for digitally obtaining images at different axial planes. Traditional image defocusing techniques such as deconvolution methods can be employed for fluorescence microscopy with various success [20]. Other related methods such as multi-focal microscopy could also be used to acquire the focal plane image, etc. [21, 22]. However, the successes of these methods depend on one or more various factors from these requirements: First, the assumption of the forward model of the image information process; second, iterative process in image reconstruction; third, additional optical components and hardware into commercial fluorescence microscopy that potentially introduces many other factors, need to be considered such as alignment, calibration, aberration, photon-efficiency, etc.. Wu et al. [23] show that DCNN can be used to propagate image from a single plane to other planes that results in the possibilities to acquire 3D volume. Previous studies have recently developed DCNN based methods for auto-focusing which result quantitative out-of-focus levels [24-26] or our proof-of-concept of image regression-based auto-focusing [27]. Recently, *Guo, et al.* [28] accelerated iterative deconvolution process for defocusing biomedical images via deep learning. In our work, we fully developed a DCNN based method that can inherently learn the optical properties governing intensity-based fluorescent wave propagation for a large out-of-focus range [-8µm 10µm] to virtually obtain the fluorescence images at the focus plane. With the advantage of not using mechanically translating or extra refocusing algorithm, this proposed end-to-end technique, together or stands alone, can improve the robustness of automatic systems such as integrated microplate microscopy or automatic mechanical scanning to acquire data in large scale. This also avoids phototoxicity and photobleaching from manual focus adjustment which are always the main concern in trade-off sample preservation, image quality, and data acquisition time.

Using the trained AF model, we have successfully predicted the focused images of the testing U2OS-AF dataset, which only contains F-actin. To check the generalizability of the trained model, we applied it directly on a completely new type of Golgi apparatus + F-actin dataset (U2OS-CPS dataset) without using transfer learning. In fact, both types of images have some similar characteristics, so the model can detect a similar set of features when using the same kernel filter. The success of the proposed AF model varies with different channels in U2OS-CPS dataset. For those images which have Golgi + f-actin channels that appear as blurry (based on stretching fibers and cell's edges), their other channels are not obvious to be distinguished. Predicting these channels using such pre-trained U2OS-AF on U2OS-AF dataset gives less accurate results. The reason is that other channels contain different spatial features that reduce the prediction accuracy of the model. For this fixable issue, transfer learning should be used if the ground truth images exist, if not, unsupervised domain adaptation [28] would be a potential solution. Using a single model to learn all targeted transformation is an easy and first approach to be implemented, but it will lead to a lower accuracy due to non-focus targeted training [16].

Recent research efforts have been developed to predict fluorescence images from unlabeled images using deep neural network, such as from bright field or phase contrast images [15, 16]. In our work, we sought to determine whether a network could generate a more complicated labeling of, for example, F-actin and Vinculin from phase contrast images, as shown in Fig. 4. Phase contrast is a common and preferable bright-field microscopy technique used to detect details of semitransparent living cells having a wide variation of refractive index common in most organelles. In fact, high-contrastive refraction of index observed from nucleus region captured by phase contrast is due to the high tendency of the nucleus to bend light passing though the objective lenses. Our phase contrast microscope produced high contrast of the cell nuclei due to index of refraction mismatch between the nuclear envelope and adjacent cytoplasm, yielding highly useful information for training of the cGAN model to predict fluorescence labels of DNA. In general, phase contrast microscopy using a high numerical aperture objective will provide great contrast and detail of membrane-bound organelles, and is expected to predict fluorescence labels of such organelles well [29, 30]. On the other hand, nano-scale structures such as cytoskeletal proteins (e.g., F-actin) and adhesion proteins (e.g., Vinculin) share similar contrast with the cytoplasm background making them poor predicted signals from phase contrast microscopy inputs to the cGAN model. However, cytoskeleton structures like F-actin are more effective inputs to the cGAN model to predict Vinculin signal, likely because F-actin and Vinculin share spatial connectedness in the cell, related to their coordinated mechanobiological function in linking focal adhesions to contractile apparatus of the cell. Vinculin are membrane-cytoskeleton proteins whose cap bind along actin filaments to either provide or prevent connections between themselves and several F-actin proteins promoting cell-cell or cell-extracellular matrix junctions [31]. Furthermore, a computational model based on vinculin - lamellipodial actins binding lifetime has been proposed recently. The bonds among them can be formed directionally and asymmetrically suggesting their high correlation in terms of spatial distributions [32].

The index of normalized, summed predicted image spatial and intensity error is an image-centric and interpretation-focused way to standardize comparisons of algorithm performance for deep learning image regression across multiple developers. For example, two other recent works predict fluorescence image outputs from ground truth brightfield images serving as input to deep learning algorithms [15, 16]. The image-wise Pearson correlation coefficient, while simple, does not highlight specific sub-regions in the image where the algorithm performs well or poorly. Further, difference images (predicted minus ground truth) are qualitative and difficult to assess, even with well-chosen colormaps. The index proposed in this study measures two competing errors with intuitive visual interpretations: pixel intensity mismatch, and area mismatch. As the tolerance for intensity mismatch normalized to the mean signal intensity of ground truth becomes larger, the area mismatch normalized to the signal image area fraction tends to become smaller. Not only would this sum of normalized errors standardize reporting across laboratories and algorithm developers, the index is also of potential utility to the end-user in two ways. The end-user can decide on the relative weights of intensity and spatial error, to be used as a target for training the algorithm on new data.

The presented methodology of fluorescence-to-fluorescence transform has a great potential to reduce repetitive and time-consuming preparation in the biological imaging process such as, sample preparation and data collection in a cost-effective manner. The advantages of using computational microscopy through deep convolutional neural networks allows a single well-tuned training process to transform fluorescence images of a certain fluorescence channel into their other fluorescence images. This will enable multiple protein predictions in image-based screens. Training the data set is a required step but it is performed only once with less pre-processing. Recent related research has demonstrated that the whole structure and organelles can be predicted from transmitted light imaging. In this work, we have extended the use of deep neural networks to predict cell's proteins from their other co-registered proteins. The proposed DCNN is an effective way in fluorescence prediction of certain proteins from other fluorescent proteins since they are invisible in bright light microscopy. Finally, the proposed DCNN will be a cost-effective tool for many biological studies of hard-defined proteins and help researchers visualize the coordination in subcellular organelles at many cell life-cycle stages and understand their biological fundamental behaviors.

In future work, inferencing mode using deep learning-based method to perform real time virtual organelle self-coding would be targeted and quantitative index would be investigated to precisely predict the complex proteins.

**Methods**

**Data preparation**

*MDA-MB-231 breast cancer cell*

Human breast cancer cell line MDA-MB-231 provided from Dr. Zaver Bhujwalla (John Hopkins School of Medicine, Baltimore, MD) was cultured on tissue treated polystyrene dish, in a standard tissue culture conditions of 37°C with 5% $CO_2$, and 100% humidity (HERAcell 150i, Thermo Fisher Scientific, Waltham, MA). Cells were fed with Dulbecco's Modified Eagle Medium (DMEM) supplemented with 10% Fetalgrow (Rocky Mountain Biologicals, Missoula, Montana) and 1% penicillin-streptomycin (Corning Inc., Corning, New York, NY). Cells were fed every two days and passaged using Trypsin (Mediatach, Inc. Manassas, VA) once they reached confluence.

MDA-MB-231 WT after passaging were seeded on 35 mm tissue culture treated dishes (CELLTREAT Scientific Products, Pepperell, MA), and followed the culture procedure provided above. After 24 hours of culturing, cells were washed with 1X Phosphate buffered saline (PBS) (Sigma-Aldrich, St. Louis, MO) to remove cell debris and fixed with 3.7% formaldehyde diluted from 16% Paraformaldehyde (Electron Microscopy Sciences, Hatfield, PA) in 15 minutes, permeabilized with 0.1%Triton-X in PBS in 5 minutes and blocked by horse serum in an hour.

Vinculin monoclonal (VLN01) antibody (Thermo Fisher Scientific, Rockford, IL) and integrin beta-1 (P5D2) antibody (Iowa University Department of Biology, Iowa, IA) was diluted in 1X PBS contained 1% Bovine Serum Albumin (BSA) to reach 2 µg/ml concentration. After one hour, cells were washed and incubated for another one hour with anti-mouse secondary immunofluorescence. Then, cells were co-stained with a solution contained 1:1000 dilution of a 2 ug/ml 4',6-diamidino-2-phenylindole (DAPI) (Life Technologies, Carlsbad, CA) and 1:1000 dilution of Phalloidins (Life Technologies Corporation, Eugene, OR) in one more hour before being washed with 1X PBS again. Cells were stored in 1X PBS at 4°C until imaging session. Fluorescence images were taken by Olympus BX60 microscope (Olympus, Tokyo, Japan), using 60×, NA 1.25 oil immersion Plan Apo objective, and a Photometrics CoolSNAP HQ2 high resolution camera (1392 x 1040 pixels, 6.45 x 6.45-µm pixels) with Meta-Morph software. MDA-MB-231 dataset contains 74 images (1392 x 1040 pixels) for training + validation and 6 images (1392 x 1040 pixels) for testing.

*Human osteosarcoma U2OS cell – Autofocusing (U2OS-AF)*

High-content screening (HCS) of U2OS cells [33] were acquired from 384-well microplate on ImageXpress Micro automated cellular imaging system with Hoechst 33342 markers and Alexa Fluor 594 phalloidin at 20x magnification, 2x binning and 2 sites per well. 32 image sets provided corresponding to 32 z-stacks with 2 µm between slices. Each image is 696 x 520 pixels in 16-bit TIF format, LZW compression. For each site, the optimal focus was found using laser auto-focusing to find the well bottom. The automated microscope was then programmed to collect a z-stack of 32 image sets covering from -32µm to 30µm of out-of-focus range.

*Human osteosarcoma U2OS cell - Cell Painting staining protocol (U2OS-CPS)*

U2OS cell (#HTB-96, ATCC) raw images that we used can be found in Ref. [18]. Cells were cultured with the density of 200 cells per will in 384-well imager with special treatment. Eight different cell organelles were adhered by different stains: nucleus (Hoechst 33342), endoplasmic reticulum (concanavalin A/AlexaFluor488 conjugate), nucleoli and cytoplasmic RNA (SYTO14 green fluorescent nucleic acid stain), Golgi apparatus and plasma membrane (wheat germ agglutinin/AlexaFluor594 conjugate, WGA), F-actin (phalloidin/AlexaFluor594 conjugate) and mitochondria (MitoTracker Deep Red). Five fluorescent channels were images at 20x magnification using Micro epifluorescent microscope with illuminating wavelength and excited wavelength as following: DAPI (387/447 nm), GFP (472/520 nm), Cy3 (531/593 nm), Texas Red (562/642 nm), Cy5 (628/692 nm). The dataset contains 3456 x 9 (folders) of each fluorescent channel (1024 x 1374 pixels). Eight folders across all channels were used for training and validation, last folder(s) were used for testing purpose.

**Training and testing data preparation**

Patch images (256 x 256 or 128 x 128) were randomly cropped from FOV to form input-output pairs for training. During that training of MDA-MB-231 breast cancer cell dataset, images were augmented with rotation and flipping to generate more features. Histogram equalization techniques were applied to enhance the image contrast (only on MDA-MB-231). All images in the datasets were pre-processed with [-1 1] normalization only. Overlapping regions of input on phase contrast or fluorescent channel were cropped randomly in horizontal and vertical directions in the training process. The predicted/tested images were divided into sub-regions with some overlap between adjacent regions to be stitched into a larger FOV based on alpha blending algorithm. Finally, the stitched predicted images were inversely normalized to the original image range.

**Conditional generative adversarial network (cGAN) implementation**

The proposed DCNN based cGAN takes one or a set of intensity images $I$ as the network input and output a single fluorescence image representing a single targeted protein. The intensity images $I$ are captured under phase contrast or fluorescence microscopy. The cGAN consists of two sub-networks (see Fig. 9), the generator G and the discriminator D. The generator G is trained to predict the proteins $\Phi_G = G(I)$ from the given input $I$. During the training process, generator G's parameters ($\theta_G$ - weights and biases of the generator) will be optimized to minimize a loss function $l$ through $N$ input-output training pairs:

$$\hat{\theta}_G = argmin_{\theta_G} \sum_{n=1}^{N} \frac{1}{N} l(G_{\theta_G}(I_n, \Phi_n)). \tag{1}$$

The generator *G* is a customized model based on the original U-Net model [34], which can adapt to efficient learning based on pixel-to-pixel transformation. A series of operations are performed, including batch-normalization (BN), nonlinear activation using ReLU/LeakyReLU (LReLU) functions, convolution (Conv2), and convolution transpose (Conv2T) layers with filters of kernel size of $k = 3$. This model contains an initial convolution layer with stride of 2 (S2), encoded-blocks (LReLU-Conv2S2-BN) and decoded-blocks (ReLU-Conv2TS2-BN), and ends with Tanh activation function.

The discriminator network *D* (contains weights and biases $\theta_G$) aims to distinguish the quality of prediction of the generator *G*. Discriminator *D* is initialized with a convolution layer stride 2, following a BN, 3 convolutional blocks (Conv2S2-BN-LReLU), one fully connected layer and sigmoid activation, filters with kernel size of $k = 5$ are used in the discriminator *D*. More details about the adversarial networks can be found in [35, 36]. The following adversarial min-max problem in term of expectation was solved to enhance the generator *G*'s performance:

$$min_{\theta_G} max_{\theta_D} E_{I,\Phi}[logD_{\theta_D}(I,\Phi) + E_I[\log(1 - D_{\theta_D}(I,G(I))], \tag{2}$$

The motivation of using discriminator *D* is to preserve the high frequency content of the predicted images. By using the conventional loss functions such as, the mean absolute error (MAE), mean square error (MSE), peak signal to noise ratio (PSNR), or structural similarity index (SSIM), the minimization of these pixel-wise loss functions will lead to solutions that have less perceptual quality. By training the generator *G* along with the discriminator *D*, the generator *G* can learn to generate the realistic images of protein prediction in case that the input-output image pairs are not strongly correlated. For that purpose, the proposed perceptual loss function $l$ is defined as a weighted sum of separate loss functions:

$$l = \lambda_1 l_{MAE} + \lambda_2 l_G + \lambda_3 l_{\theta_G}, \tag{3}$$

where

$$l_{MAE} = \frac{1}{W \times H} \left\| |\Phi| - |G_{\theta_G}(I)| \right\|, \quad (4)$$

$$l_G = -log D_{\theta_D}(I, G(I)), \quad (5)$$

$$l_{\theta_G} = \|\theta_G\|, \quad (6)$$

where $\|.\|$ denotes the $L_1$-norm, $(\lambda_1, \lambda_2, \lambda_3)$ are hyper parameters that control the relative weights of each loss components and were choose as $(\lambda_1 = 0.99, \lambda_2 = 0.01 \text{ and } \lambda_3 = 0.001)$, $W \times H$ is the input image size. Adaptive momentum (Adam) optimizer was used to optimize the loss function with a learning rate of 2e-4 for both the generator and the discriminator models. These models were implemented using Tensorflow framework on GPU RTX 2070 16Gb RAM Intel Core i7 and GPU Titan Xp 8Gb RAM Intel Core i7, and the models were selected based on the best performance on validation dataset.

**Quantitative index of predicted image error**

A quantitative index of predicted image error versus ground truth was defined as the sum of normalized pixel-wise intensity and spatial errors, computed over a range of tolerances of the 8-bit absolute difference error. The rationale for this index was that epifluorescence images of cell labels are qualitative in pixel intensity due to photobleaching and differences in experimental preparation, microscope instrument parameters and camera settings. Therefore, small differences between pixel intensity values of predicted images and ground truth (intensity errors) are less likely to produce misinterpretations than predicted signal where there is no true signal, or lack of predicted signal where there is true signal (spatial errors). Segmentation of absolute difference error maps so that error pixels above a certain tolerance are bright green highlights both errors. The sum of these two error terms, weighting each equally, is minimized at a single error tolerance level.

$$TL = \min(\beta_1 \times IE(O, G, i) + \beta_2 \times SE(O, G, i)), \quad (7)$$

where $TL$ is the tolerance level; $O, G$ are the input/predicted image and ground truth, respectively; $i$ is the threshold levels crossing from 0 to 99% of bit-depth range of image; $\beta_1, \beta_2$ are weights; $IE$ is intensity error function that measures the mean absolute error of two images below all threshold levels, and be normalized by maximum of bit-depth. $SE$ is binary segmented error function based on threshold that measures the area fraction between segmented areas on two images above all threshold levels.

**Acknowledgements**

The authors thank Dr. Zaver Bhujwalla and Byung-Min Chung for the gift of MDA-MB-231 breast cancer cell lines and NVIDIA Corporation for sponsoring a GeForce Titan Xp through the GPU Grant Program.

**Author Contributions**

T.N. designed, conceived and built the prototype; T.N., V.B., A.T., and V.L performed the experiments and processed the data; T.N. built the deep neural network; T.N., V.B. and A.T. trained neural network models; T.N., V.B. and A.T prepared data analysis and statistics; V.L. and

**Figure legends**

**Figure 1.** Virtual fluorescent imaging pipeline for cell microscopy with all models implemented. MDA-MB-231 cells are imaged to collect co-registered phase contrast, DAPI, F-actin and vinculin. PhC-Fluo models were trained to predict fluorescent images from phase contrast (either DAPI, or Vinculin). While Fluo-Fluo models (using MDA-MB-231 in Fluo-Fluo model 1 and U2OS-CPS in Fluo-Fluo models 2, 3, 4) will predict florescent images of other fluorescence images of same cells. Out of focus images were fed in the autofocusing model (AF model) (center) as a pre-processing step to refocus the images before being reused in training and testing processes in Fluo-Fluo models.

**Figure 2.** Deep-learning-based autofocusing fluorescent prediction. Autofocusing model is tested on U2OS-AF dataset. First row is the full FOV of a testing image at different z depths. Second row is the zoom-in areas as the inputs. Third row is the zoom-in area prediction, fourth row is the zoom-in area of the ground truth which is the same for all z-depth. First left column and bottom row with bounding boxes are extra zoom-in areas marked by colors and numbers.

**Figure 3.** Deep-learning-based autofocusing fluorescent prediction on new type dataset. Direct prediction of out-focus Golgi apparatus + F-actin in U2OS-CPS dataset using the pre-trained model trained on F-actin channel in U2OS-AF dataset.

**Figure 4.** Deep-learning-based fluorescent prediction of MDA-MB-231 dataset on testing data. First two rows, prediction of DAPI/Hoechst, Vinculin, respectively, from phase contrast. Third row is prediction of Vinculin from F-actin.

**Figure 5.** Comparison between PhC-Fluo 2 and Fluo-Fluo 1 on Vinculin prediction on a sample in testing dataset: (A) full FOV and zoom-in regions of ($A_1$) phase contrast image as the model's input, ($A_2$) Vinculin prediction of PhC-Fluo 2 model, ($A_3$) Vinculin expression distribution of interest (at cell's edges) used intensity-based threshold (19% of 255 bit-depth); (B) full FOV and zoom-in regions of ($B_1$) F-actin image as the model's input, ($B_2$) Vinculin prediction of Fluo-Fluo 1 model, ($B_3$) Vinculin expression distribution of interest (at cell's edges) used intensity-based threshold (19% of 255 bit-depth); (C) corresponding ground truth of vinculin; (D) toleration level v/s bit-depth threshold computed on 8 testing images (left to right), first row is the intensity error (IE) and area fraction (intensity-based segmentation) error (SE) of both PhC_Fluo 2 and Fluo-Fluo 1 models, second row is the summation of IE and SE of both PhC_Fluo 2 and Fluo-Fluo 1 models.

**Figure 6.** Deep-learning-based fluorescent prediction of U2OS-CPS cell on testing data: (A, B) Nucleoli + Cytoplasmic RNA and Golgi apparatus + F-actin are used as the input of the cGAN model. ($C_1$, $D_1$, and $E_1$) are the targeted fluorescent images corresponding to (DAPI/Hoechst, Mitochondria, and Endoplasmic Reticulum, respectively) as ground truth and their cGAN corresponding prediction ($C_2$, $D_2$, and $E_2$, respectively). ($F_1$ and $F_2$) merged-channel images for ground truth and prediction, respectively. Scale bar is 25 μm.

**Figure 7.** Pearson product-moment correlation coefficient of each feature measurement across 96 images on testing data between original (5) channels and hybrid-virtual (2+3) channels. "N" marks the correlation coefficients of features measured only on 2-channel input images in both cases (Golgi apparatus + F-actin) which results perfect correlations. "Inf" marks un-resolved correlation due to 0-division in measurements. Those features are distributed across 3 compartments: Cell, nuclei, and cytoplasm [18] (see Supplementary Method for feature's organization and Supplementary Figure 4 for full feature measurements).

**Figure 8.** Histogram of Pearson product-moment correlation coefficient across 96 images of each feature measurement.

**Figure 9.** Conditional generative adversarial network (cGAN) for fluorescent image prediction. This figure only shows one of the models. PhGolgi + F-actin are the inputs to predict endoplasmic reticulum as the targeted fluorescent images. Generator Network *G* contains an initial convolution layer with stride of 2 (S2), encoded-blocks (LReLU-Conv2S2-BN) and decoded-blocks (ReLU-Conv2TS2-BN), ends with Tanh activation and uses skip connections. Generator Network transforms the input images and results in predicting fluorescent images. Discriminator Network *D* is initialized with a convolution layer stride 2, following a BN, 3 convolutional blocks (Conv2S2-BN-LReLU), one fully connected layer and sigmoid activation. Discriminator *D* outputs a score of how likely the input of a group of images is good or bad. The input of Discriminator *D* was formed as a conditional input by concatenating the predicted image or ground truth with generator *G*'s input.

**Figures**

**Figure 1**

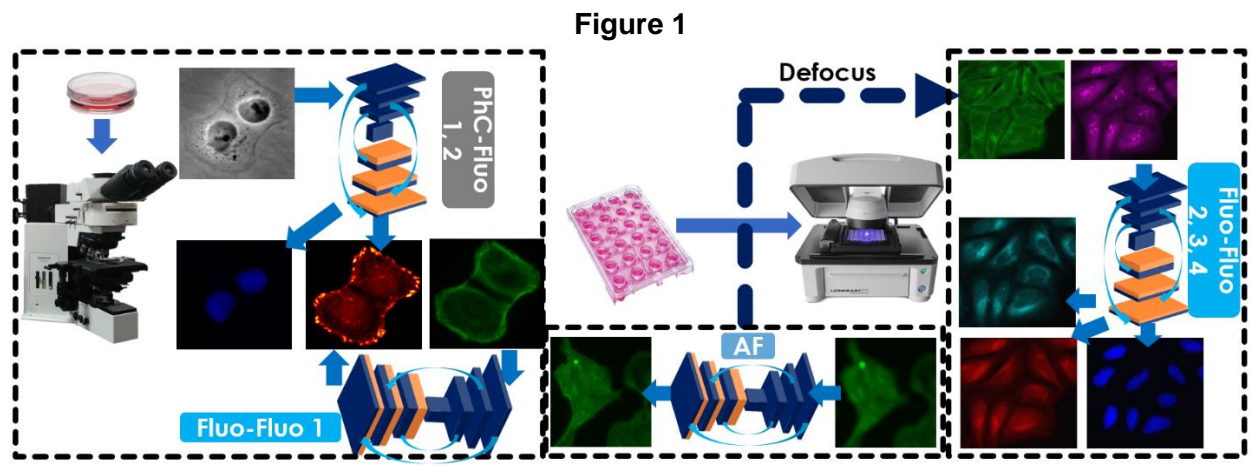

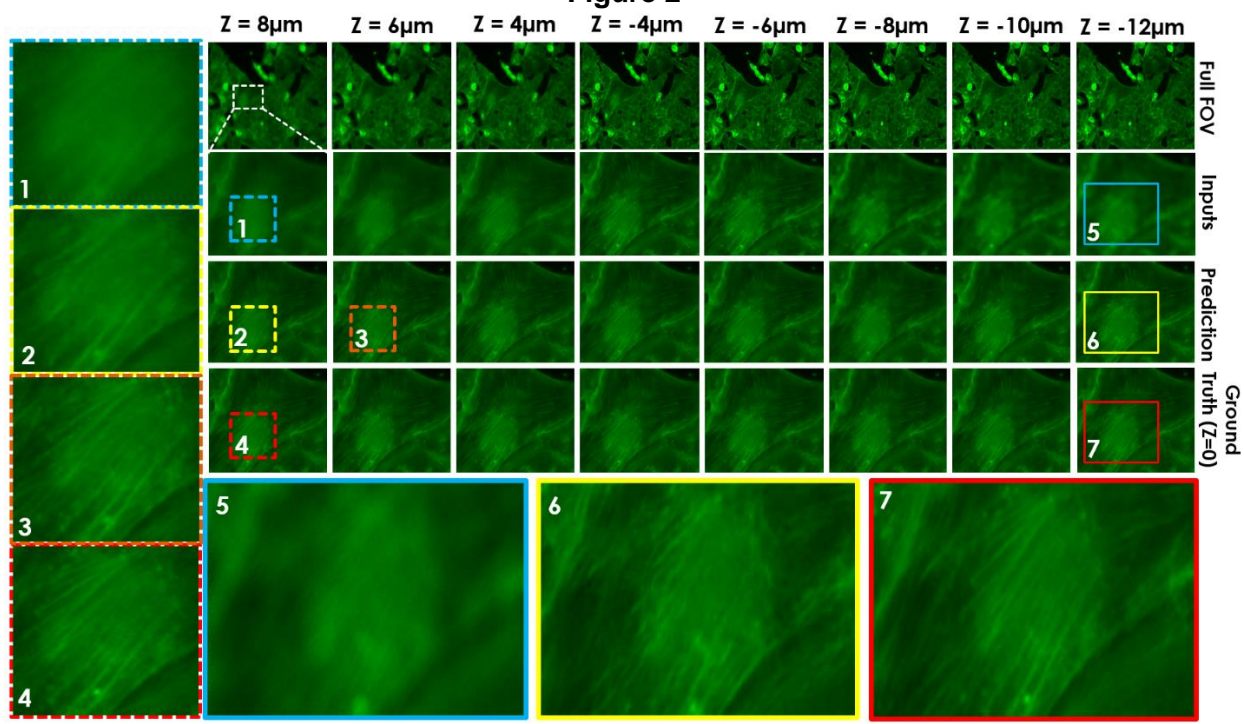

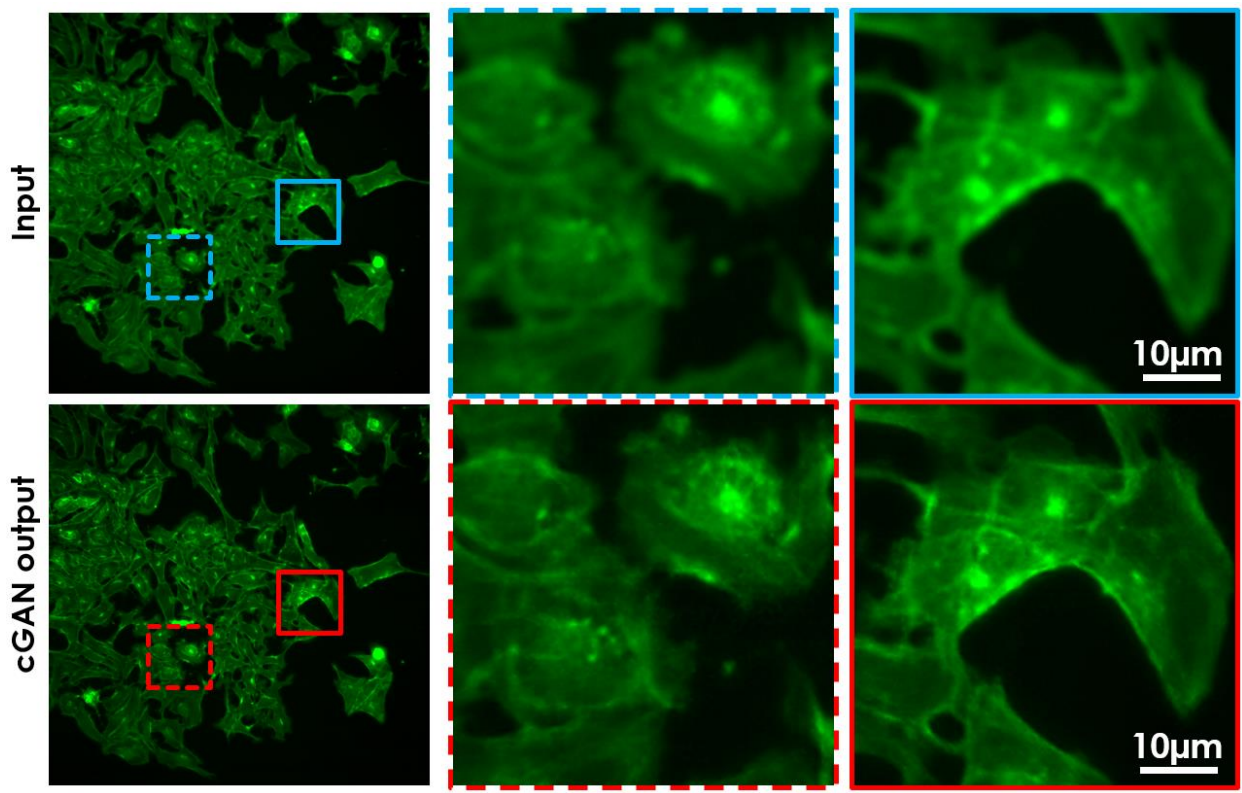

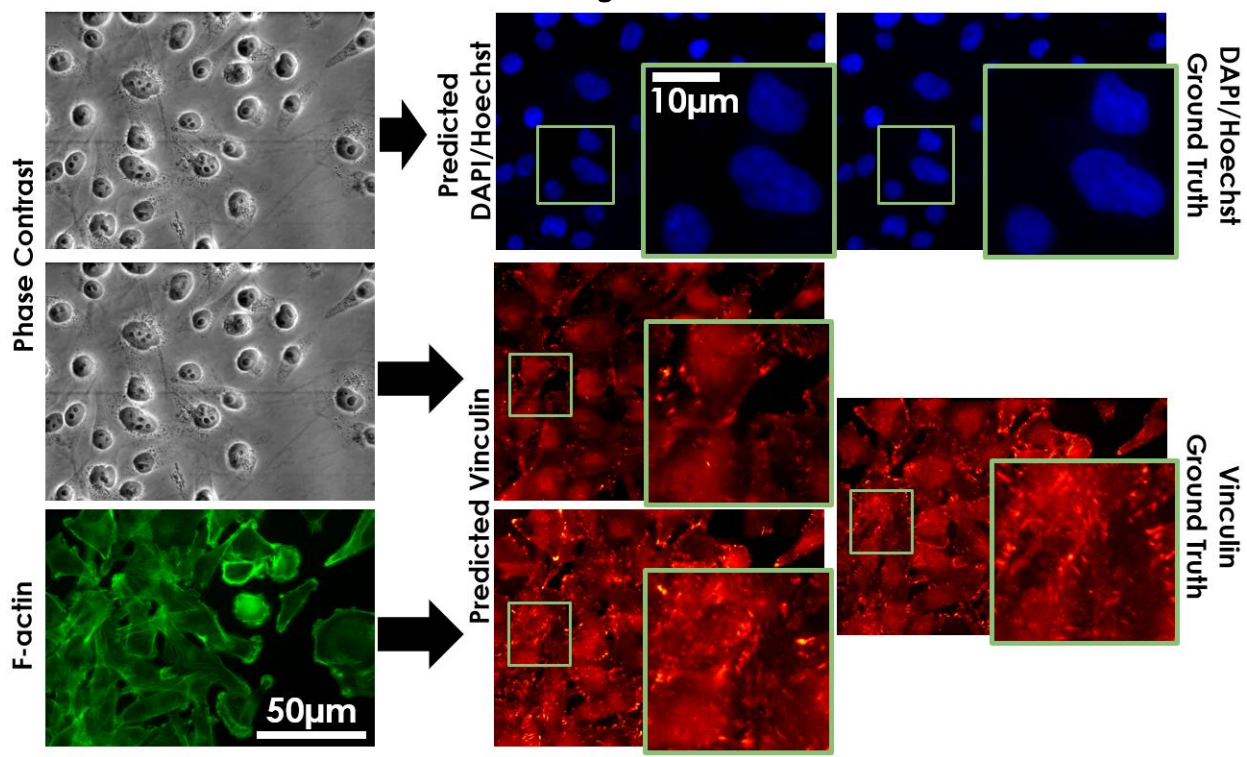

Figure 4

**Figure 5**

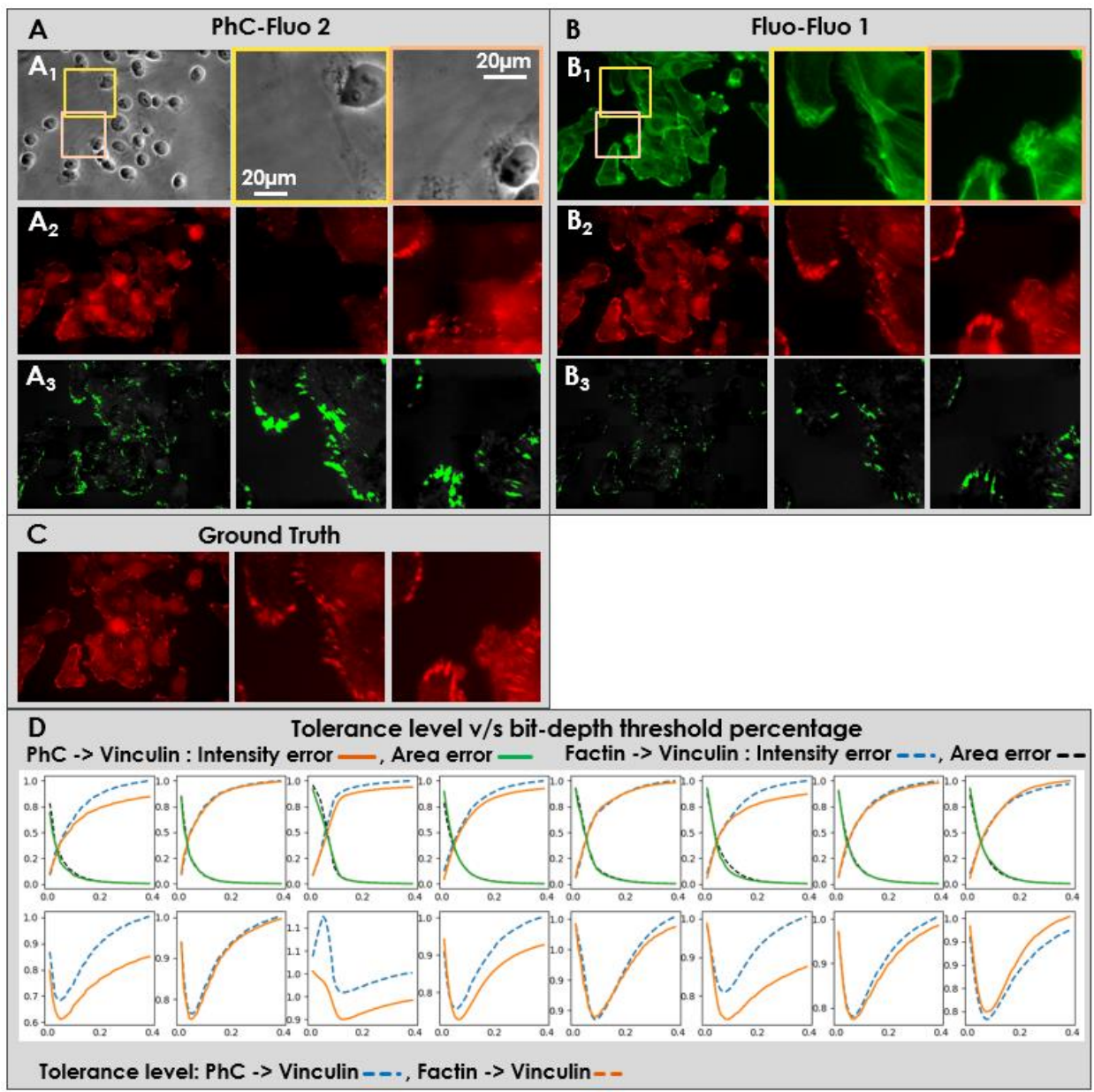

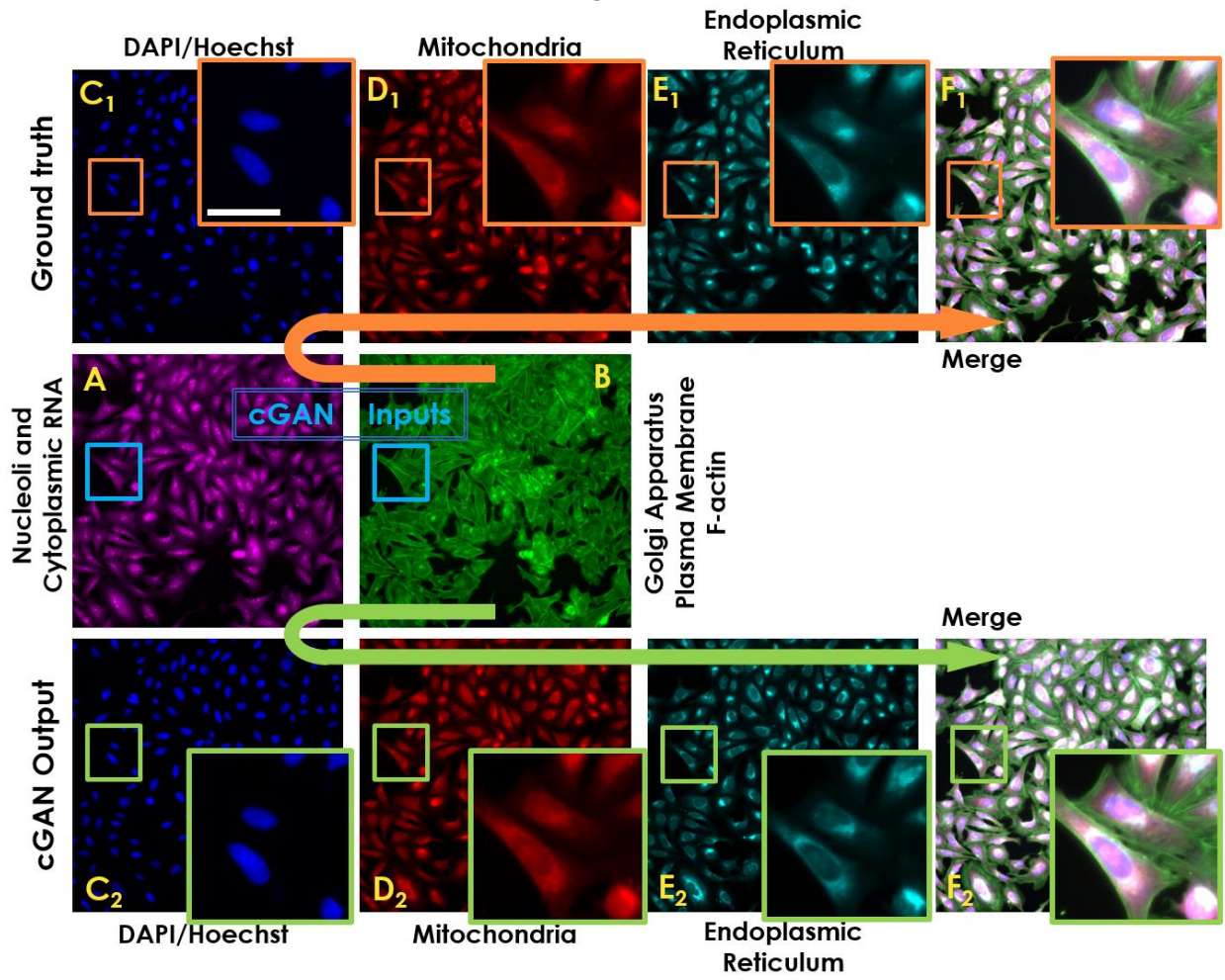

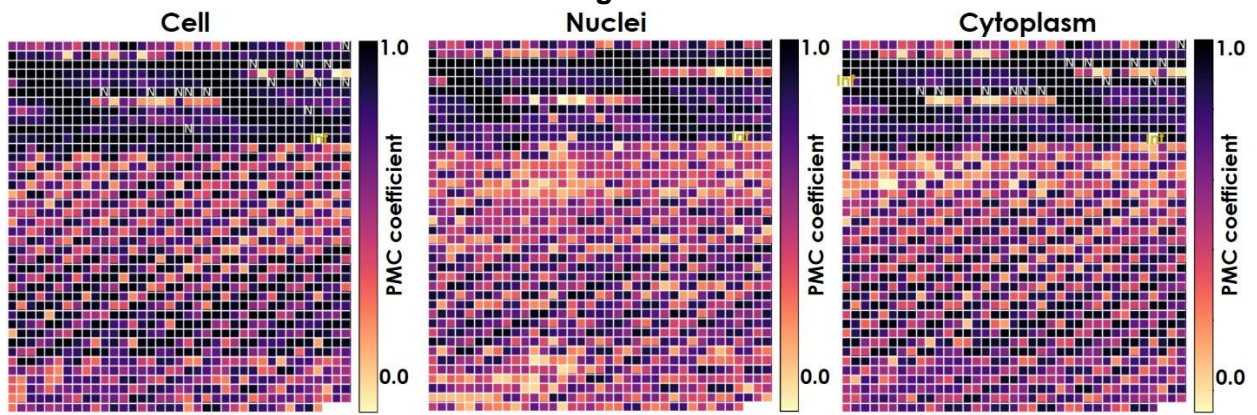

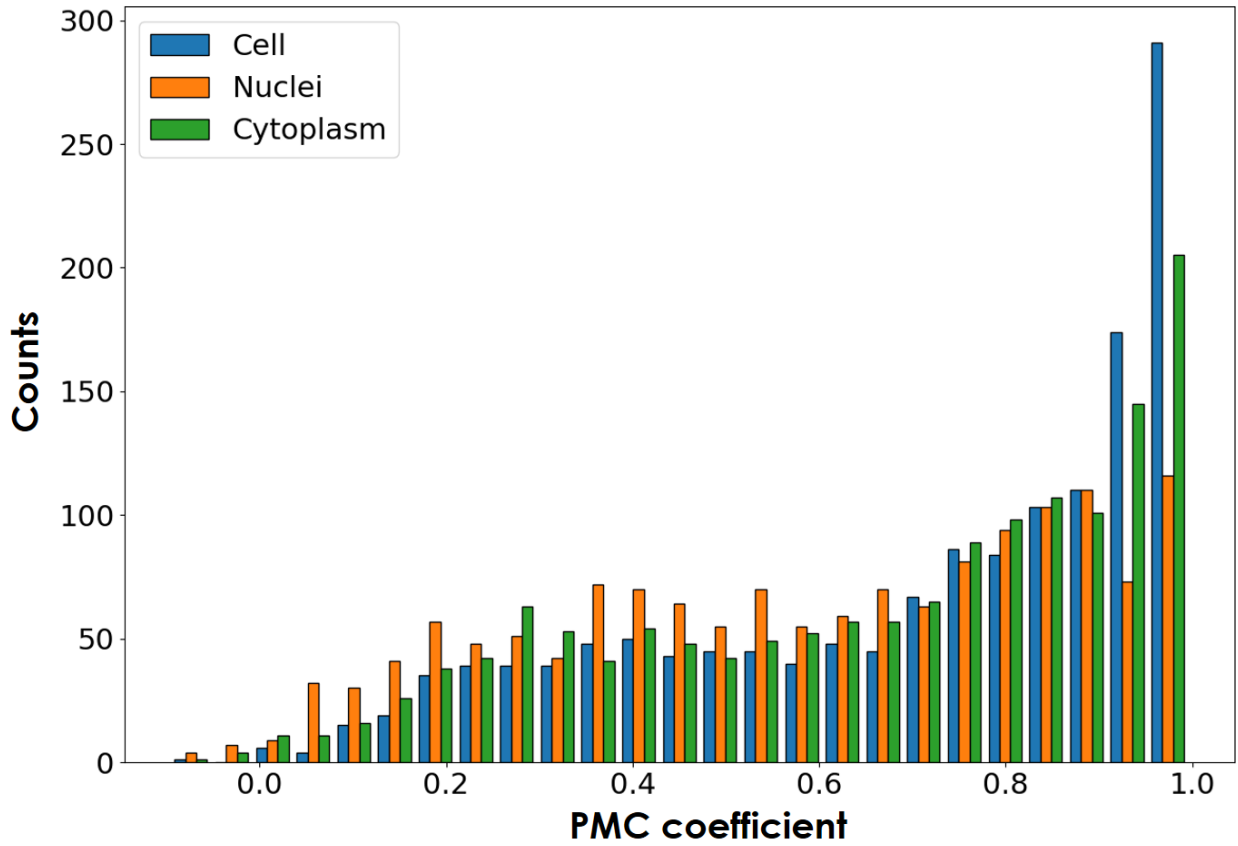

Figure 8

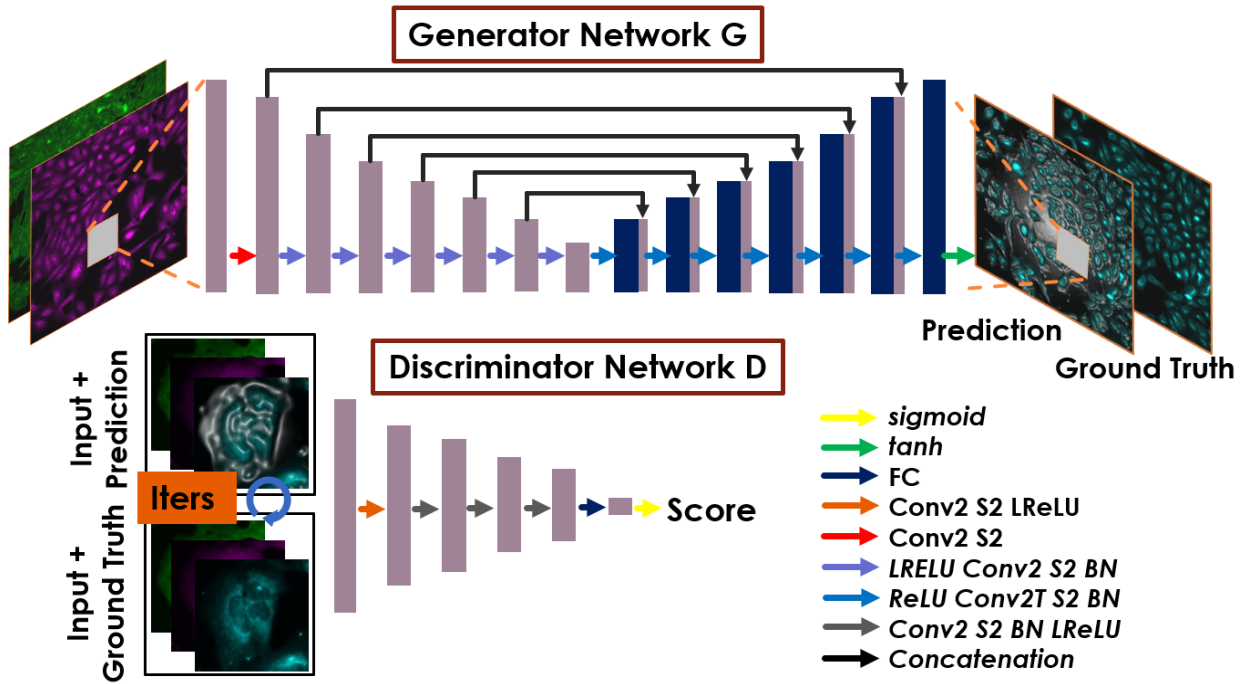

Figure 9